\documentclass[nofootinbib,prd,aps,superscriptaddress,preprintnumbers,twocolumn,showpacs]{revtex4-2}
\usepackage[dvipdfmx]{graphicx}
\usepackage{amsmath,amssymb,amsthm,bm,bigdelim,multirow}
\usepackage{color}

\usepackage{booktabs}
\usepackage{hyperref}
\usepackage{dsfont}
\usepackage{here}

\newcommand{\kslash}{k\kern-1ex /}
\newcommand{\pslash}{p\kern-1ex /}
\newcommand{\qslash}{q\kern-1ex /}
\newcommand{\lslash}{l\kern-1ex /}
\newcommand{\sslash}{s\kern-1ex /}
\newcommand{\Dslash}{D\kern-1.2ex /}

\newcommand{\beqa}{\begin{eqnarray}}
\newcommand{\eeqa}{\end{eqnarray}}
\newcommand{\Tr}{{\rm Tr}}

\newcommand{\bd}{\begin{description}}
\newcommand{\ed}{\end{description}}
\newcommand{\la}{\langle}
\newcommand{\ra}{\rangle}
\newcommand{\ben}{\begin{eqnarray}}
\newcommand{\een}{\end{eqnarray}}
\newcommand{\nn}{\nonumber}
\def\lsim{\raise0.3ex\hbox{$<$\kern-0.75em\raise-1.1ex\hbox{$\sim$}}}
\def\gsim{\raise0.3ex\hbox{$>$\kern-0.75em\raise-1.1ex\hbox{$\sim$}}}
\def\simgt{\rlap{\lower 3.5 pt\hbox{$\mathchar \sim$}}\raise 2.0pt \hbox {$>$}}
\def\simlt{\rlap{\lower 3.5 pt\hbox{$\mathchar \sim$}}\raise 2.0pt \hbox {$<$}}

\begin{document}

\title{Tensor renormalization group approach to (1+1)-dimensional SU(2) principal chiral model at finite density}

\author{Xiao Luo,}
	\affiliation{Graduate School of Pure and Applied Sciences, University of Tsukuba, Tsukuba, Ibaraki
    305-8571, Japan}

  	\author{Yoshinobu Kuramashi}
  	\affiliation{Center for Computational Sciences, University of Tsukuba, Tsukuba, Ibaraki
    305-8577, Japan}

\begin{abstract}
  We apply the tensor renormalization group method to the (1+1)-dimensional SU(2) principal chiral model at finite chemical potential with the use of the Gauss-Legendre quadrature to discretize the SU(2) Lie group. The internal energy at vanishing chemical potential $\mu=0$ shows good consistency with the prediction of the strong and weak coupling expansions. This indicates an effectiveness of the Gauss-Legendre quadrature for the partitioning of the SU(2) Lie group. In the finite density region with $\mu\ne 0$ at the strong coupling we observe the Silver-Blaze phenomenon for the number density. 
\end{abstract}
    
\date{\today}

\maketitle

\section{Introduction}
\label{sec:intro}

The tensor renormalization group (TRG) method \footnote{In this paper, the ``TRG method" or the ``TRG approach" refers to not only the original numerical algorithm proposed by Levin and Nave \cite{Levin:2006jai} but also its extensions \cite{PhysRevB.86.045139,Shimizu:2014uva,Sakai:2017jwp,Adachi:2019paf,Kadoh:2019kqk,Akiyama:2020soe,PhysRevB.105.L060402,Akiyama:2022pse}.} is a deterministic numerical algorithm with no sign problem, whose basic idea was proposed in the field of condensed matter physics in 2007~\cite{Levin:2006jai}. Since the original algorithm was designed to study only two-dimensional (2$d$) classical spin systems, it was necessary to develop new algorithms and calculational techniques to apply the TRG method  to particle physics, where we have to treat various theories consisting of the scalar, gauge, and fermion fields on the (3+1)$d$ space-time. Our first task was to develop efficient methods to treat the scalar, gauge, and fermion fields verifying the following expected advantages of the TRG method using the lower-dimensional models: (i) no sign problem \cite{Shimizu:2014uva,Shimizu:2014fsa,Shimizu:2017onf,Takeda:2014vwa,Kadoh:2018hqq,Kadoh:2019ube,Kuramashi:2019cgs}, (ii) logarithmic computational cost on the system size, (iii) direct manipulation of the Grassmann variables \cite{Shimizu:2014uva,Sakai:2017jwp,Yoshimura:2017jpk}, (iv) evaluation of the partition function or the path-integral itself. Recently, we have gradually moved on to the next stage to study various (3+1)$d$ models with the TRG method. So far we have analyzed the phase transitions of the Ising model~\cite{Akiyama:2019xzy}, the complex $\phi^4$ theory at finite density~\cite{Akiyama:2020ntf}, the real $\phi^4$ theory~\cite{Akiyama:2021zhf}, the Nambu$-$Jona-Lasinio (NJL) model at high density and very low temperature~\cite{Akiyama:2020soe}, and $\mathds{Z}_2$ gauge-Higgs model at finite density~\cite{Akiyama:2022eip}. The next target would be the (3+1)$d$ non-Abelian gauge theories, especially, with the SU(2) and SU(3) gauge groups. 

A difficulty in treating the non-Abelian gauge theories with the TRG method stems from the discretization for the parameter space of the Lie group. In the previous studies of the scalar field theories with the TRG method, the continuous degrees of freedom are successfully discretized with the Gauss quadrature~\cite{Kadoh:2018hqq,Kadoh:2018tis,Kadoh:2019ube,Akiyama:2020ntf,Akiyama:2021zhf}. It also works well for the 2$d$ U(1) gauge theory with a $\theta$ term~\cite{Kuramashi:2019cgs}. Based on these experiences it is worth to apply the Gauss quadrature to the SU(2) case\footnote{Other possible approaches to non-abelian gauge theories are proposed in Refs.~\cite{Liu:2013nsa,Fukuma:2021cni,Hirasawa:2021qvh,Kuwahara:2022ubg}.}. Before exploring the (3+1)$d$ SU(2) and SU(3) gauge theories, it would be better to check the efficiency of the discretization method for the non-Abelian group employing a lower-dimensional model. In this paper we investigate the (1+1)$d$ SU(2) principal chiral model (PCM) at finite density. We examine the efficiency of the Gauss quadrature by comparing the results of the internal energy at zero density to those of the strong and weak coupling expansions~\cite{Rossi:1993zc,Guha:1983ib} over the wide range of the coupling constant. Once the efficiency of the Gauss quadrature is confirmed, we investigate the $\mu$ dependence of the number density at finite density changing the chemical potential $\mu$ systematically. 

This paper is organized as follows. In Sec.~\ref{sec:method}, we define the (1+1)-dimensional SU(2) PCM at finite density on the lattice and explain how to construct its tensor network representation. We present the results for the internal energies at $\mu=0$ and the number density at $\mu\ne 0$ in Sec.~\ref{sec:results}.  
Section~\ref{sec:summary} is devoted to summary and outlook.
In the appendix we show our results for the (1+1)$d$ O(3) nonlinear sigma model at finite density in comparison with those obtained by the dual lattice simulation~\cite{Bruckmann:2016txt}. The consistency between them provides validity of our method against the sign problem in the finite density region.

\section{Formulation and numerical algorithm}
\label{sec:method}

\subsection{(1+1)-dimensional SU(2) principal chiral model at finite density}
\label{subsec:action}

We consider the partition function of the SU(2) PCM at finite density on an isotropic hypercubic lattice $\Lambda_{1+1}=\{(n_1,n_2)\ \vert n_{1,2}=1,\dots,L\}$ whose volume is equal to $V=L\times L$. The lattice spacing $a$ is set to $a=1$ without loss of generality. The SU(2) matrix $U(n)$ reside on the sites $n$ and satisfies the periodic boundary conditions $U(n+{\hat \nu}L)=U(n)$ ($\nu=1,2$). Following Ref.~\cite{Gattringer:2017hhn} we take the lattice action $S$ defined by
\begin{widetext}
\begin{align}
\label{eq:action}
S=-\beta N \sum_{n\in\Lambda_{1+1},\nu} \left\{{\Tr}\left[ e^{\delta_{\nu,2}\sigma_3\frac{\mu_1+\mu_2}{2}}U(n)e^{\delta_{\nu,2}\sigma_3\frac{\mu_1-\mu_2}{2}}U^\dagger(n+\hat{\nu})\right]+{\Tr}\left[ e^{-\delta_{\nu,2}\sigma_3\frac{\mu_1-\mu_2}{2}}U^\dagger (n)e^{-\delta_{\nu,2}\sigma_3\frac{\mu_1+\mu_2}{2}}U(n+\hat{\nu})\right]\right\},
\end{align}
\end{widetext}
where $\mu_{1,2}$ are the chemical potentials coupled to two Noether charges and $N=2$ for the SU(2) group in this work. $\sigma_3$ is the third generator in the SU(2) group. Note that this model suffers from the complex action problem in case of $\mu_{1,2}\ne 0$. The partition function $Z$ is given by
\ben
\label{eq:partitionfunction}
Z=\int{\cal D}[U]e^{-S},
\een
where ${\cal D}[U]$ is the SU(2) Haar measure, whose expression is given later.

\subsection{Tensor network representation of the model}
\label{subsec:tn-rep}

The SU(2) matrix in PCM can be expressed as 
\begin{equation}
	U = s_0 \sigma_0 +  is_j \sigma_j, \quad j=1,2,3,
\end{equation}
where $\sigma_i$ are the generators in the SU(2) group, and $\sigma_0$ is an identity matrix. Moreover, this follows $U^\dagger U = 1$, which means the SU(2) PCM can be represented by expression of O(4) $\sigma$ model
\begin{equation}
	\label{U:representation}
	\begin{array}{c}
		 \bm{s}^T(\Omega) = (\cos\psi, \sin\psi \cos\theta, \sin\psi \sin\theta \cos\phi, \sin\psi \sin\theta \sin\phi) \\
		 \Omega=(\psi,\theta,\phi) \quad, ~\psi,\theta \in (0,\pi],~\phi \in (0,2\pi]. 
	\end{array}
\end{equation}
After expanding $e^{t\sigma_i}$ to $\cosh t + \sigma_i \sinh t$,
we obtain the representation of the action (\ref{eq:action}) as follows:
\begin{equation}
	\label{eq:actiono4}
	S = -2N^2 \beta \sum_{n\in\Lambda_{1+1},\nu} s_i(n) D_{ij}(\mu_1,\mu_2;\hat{\nu}) s_j(n+\hat{\nu}),
\end{equation}
where $D_{ij}(\mu_1,\mu_2;\hat{\nu})$ is a 4$\times$4 matrix expressed as
\begin{widetext}
\begin{align}
	\label{eq:finitdensitymatrix}
	D_{ij}(\mu_1,\mu_2;\hat{\nu}) = \left(\begin{array}{rrrr}
		\cosh(\delta_{\nu,2} \mu_1) & ~ & ~ & -i\sinh(\delta_{\nu,2} \mu_1) \\
		~ & \cosh(\delta_{\nu,2} \mu_2) & i\sinh(\delta_{\nu,2} \mu_2) & ~ \\
		~ & -i\sinh(\delta_{\nu,2} \mu_2)  & \cosh(\delta_{\nu,2} \mu_2) & ~ \\
		i\sinh(\delta_{\nu,2} \mu_1) & ~ & ~ & \cosh(\delta_{\nu,2} \mu_1)
	\end{array}\right) 
\end{align}
\end{widetext}
with the use of the commutation relations of the SU(2) generators.
Note that the action becomes complex in case of the finite density. 
The partition function and its measure are written as
\begin{align}
	\label{eq:partitionfunction2}
	Z &= \int {\cal D}\Omega \prod_{n,\nu} e^{2N^2 \beta  s_i(\Omega_n) D_{ij}(\mu_1,\mu_2;\hat{\nu}) s_j(\Omega_{n+\hat{\nu})}}, \\
	{\cal D}\Omega &= \prod_{p=1}^{V} \frac{1}{2\pi^2} \sin^2(\psi_p) \sin(\theta_p) d\psi_p d\theta_p d\phi_p~.
\end{align}
We discretize the integration (\ref{eq:partitionfunction2}) with the Gauss-Legendre quadrature~\cite{Kuramashi:2019cgs,Akiyama:2020ntf} after changing the integration variables:
\ben
&-1 \le \alpha=\frac{1}{\pi}\left(2\psi-\pi \right)\le 1, &\\
&-1 \le \beta=\frac{1}{\pi}\left(2\theta-\pi \right)\le 1, &\\
&-1 \le \gamma=\frac{1}{\pi}\left(\phi-\pi \right)\le 1. &
\een
We obtain
\begin{widetext}
\begin{align}
	Z = \sum_{ \{\Omega_1\},\cdots,\{\Omega_V\}} \left( \prod_{n=1}^{V} \frac{\pi}{8}  \sin^2(\psi(\alpha_{a_n})) \sin(\theta(\beta_{b_n})) w_{a_n} w_{b_n} w_{c_n} \right) \prod_{\nu} M_{\Omega_n,\Omega_{n+\hat{\nu}}}
\end{align}
\end{widetext}
with $\Omega_n=(\psi(\alpha_{a_n}),\theta(\beta_{b_n}),\phi(\gamma_{c_n}))\equiv (a_n,b_n,c_n)$, where $\alpha_{a_n}$, $\beta_{b_n}$, $\gamma_{c_n}$ are $a$-, $b$-, $c$-th roots of the $K$-th Legendre polynomial $P_{K}(s)$ on the site $n$, respectively. $\sum_{ \{\Omega_n\}}$ denotes $\sum_{a_n=1}^{K}\sum_{b_n=1}^{K}\sum_{c_n=1}^{K}$.
$M$ is a 6-legs tensor defined by
\begin{widetext}
\begin{align}
	M_{a_n,b_n,c_n,a_{n+\hat{\nu}}, b_{n+\hat{\nu}}, c_{n+\hat{\nu}}} 
  = \exp\left\{ 2N^2\beta s_i(a_n,b_n,c_n) D_{ij}(\mu_1,\mu_2;\hat{\nu}) s_j(a_{n+\hat{\nu}}, b_{n+\hat{\nu}}, c_{n+\hat{\nu}}) \right\}~.
\end{align}
\end{widetext}
The weight factor $w$ of the Gauss-Legendre quadrature is defined as
\begin{widetext}
\begin{align}
	w_{a_n} = \frac{2(1-{\alpha_{a_n}}^2)}{K^2P^2_{K-1}({\alpha_{a_n}})},\quad
	w_{b_n} = \frac{2(1-{\beta_{b_n}}^2)}{K^2P^2_{K-1}({\beta_{b_n}})},\quad 
	w_{c_n} = \frac{2(1-{\gamma_{c_n}}^2)}{K^2P^2_{K-1}({\gamma_{c_n}})}.
\end{align}
\end{widetext}
After performing the singular value decomposition (SVD) on $M$:
\begin{widetext}
\begin{equation}
M_{a_n,b_n,c_n,a_{n+\hat{\nu}}, b_{n+\hat{\nu}}, c_{n+\hat{\nu}}} 
\simeq \sum_{i_n=1}^{D_\text{cut}} U_{a_n,b_n,c_n, i_n} (\nu) \sigma_{i_n}(\nu) V^\dagger_{i_n,a_{n+\hat{\nu}}, b_{n+\hat{\nu}}, c_{n+\hat{\nu}}} (\nu),
\end{equation}
\end{widetext}
we can obtain the tensor network representation of the SU(2) PCM on the site $n\in\Lambda_{1+1}$
\begin{widetext}
\begin{align}
	T_{i_n, j_n, k_n, l_n} &= \frac{\pi}{8} \sqrt{\sigma_{i_n}(1) \sigma_{j_n}(1) \sigma_{k_n}(2) \sigma_{l_n}(2) } \sum_{a_n, b_n, c_n} w_{a_n} w_{b_n} w_{c_n} \sin^2(\psi_{a_n}) \sin(\theta_{b_n})\nonumber \\
	&\quad \times V^\dagger_{i_n,a_n,b_n,c_n} (1) U_{a_n,b_n,c_n, j_n} (1) V^\dagger_{k_n,a_n,b_n,c_n} (2) U_{a_n,b_n,c_n, l_n} (2) , 
\end{align}
\end{widetext}
where $D_{\text{cut}}$ is the bond dimension of tensor $T$, which controls the numerical precision in the TRG method. The tensor network representation of partition function is given by
\begin{equation}
  Z \simeq \sum_{i_0 j_0 k_0 l_0 \cdots} \prod_{n \in \Lambda_{1+1}} T_{i_n j_n k_n l_n} = \Tr \left[T \cdots T\right]~.
  \label{eq:Z_TN}
\end{equation}
We employ the higher order tensor renormalization group (HOTRG) algorithm~\cite{PhysRevB.86.045139} to evaluate $Z$.

In this work we calculate two physical quantities: internal energy and number desity. The operator of the internal energy is defined by the average of all links between the nearest sites:
\begin{align}
	\label{eq:internalenergy}
	E =& 1 -  \frac{1}{Vd}\sum_{n,{\nu}} \left\langle  s_i(n) D_{ij}(\mu_1,\mu_2,\hat{\nu}) s_j(n+\hat{\nu}) \right\rangle \nonumber \\
	=& 1 -  \frac{1}{d}\sum_{{\nu}} \left\langle  s_i(0) D_{ij}(\mu_1,\mu_2,\hat{\nu}) s_j(\hat{\nu}) \right\rangle.
\end{align}
Note that the matrix $D$ is reduced to be the identity matrix in case of zero density.
The internal energy can be obtained by the numerical differentiation of the free energy in terms of $\beta$ or the impure tensor method~\cite{Gu_2008,Kadoh:2018tis}. Since the numerical accuracy of the former method depends on the interval of $\beta$, we employ the latter one to calculate the internal energy. With the use of two types of the impure tensors
\begin{widetext}
\begin{align}
	\label{eq:internalenergytensor1}
	\tilde{T}_{i_0, j_0, k_0, l_0,\lambda} &= \frac{\pi}{8} \sqrt{\sigma_{i_0}(1) \sigma_{j_0}(1) \sigma_{k_0}(2) \sigma_{l_0}(2) } \sum_{a_0, b_0, c_0} w_{a_0} w_{b_0} w_{c_0}  \sin^2(\psi_{a_0}) \sin(\theta_{b_0})  \nonumber \\
	&\quad \times s_\lambda(a_0,b_0,c_0) V^\dagger_{i_0,a_0,b_0,c_0} (1) U_{a_0,b_0,c_0, j_0} (1) V^\dagger_{k_0,a_0,b_0,c_0} (2) U_{a_0,b_0,c_0, l_0} (2), \\
	\label{eq:internalenergytensor2}
	\tilde{T}_{i_{\hat{\nu}}, j_{\hat{\nu}}, k_{\hat{\nu}}, l_{\hat{\nu}}, \lambda} &= \frac{\pi}{8} \sqrt{\sigma_{i_{\hat{\nu}}}(1) \sigma_{j_{\hat{\nu}}}(1) \sigma_{k_{\hat{\nu}}}(2) \sigma_{l_{\hat{\nu}}}(2) } \sum_{a_{\hat{\nu}}, b_{\hat{\nu}}, c_{\hat{\nu}}} w_{a_{\hat{\nu}}} w_{b_{\hat{\nu}}} w_{c_{\hat{\nu}}}  \sin^2(\psi_{a_{\hat{\nu}}}) \sin(\theta_{b_{\hat{\nu}}}) \nonumber\\ 
	&\quad \times D_{\lambda\gamma}(\mu_1,\mu_2,\hat{\nu}) s_\gamma(a_{\hat{\nu}},b_{\hat{\nu}},c_{\hat{\nu}})  V^\dagger_{i_{\hat{\nu}},a_{\hat{\nu}},b_{\hat{\nu}},c_{\hat{\nu}}} (1) U_{a_{\hat{\nu}},b_{\hat{\nu}},c_{\hat{\nu}}, j_{\hat{\nu}}} (1) V^\dagger_{k_{\hat{\nu}},a_{\hat{\nu}},b_{\hat{\nu}},c_{\hat{\nu}}} (2) U_{a_{\hat{\nu}},b_{\hat{\nu}},c_{\hat{\nu}}, l_{\hat{\nu}}} (2), 
\end{align}
\end{widetext}
the internal energy (\ref{eq:internalenergy}) is obtained by the following tensor product:
\ben
&&\left\langle  s_\lambda(0) D_{\lambda\gamma}(\mu_1,\mu_2,\hat{\nu}) s_\gamma(\hat{\nu}) \right\rangle  \nn\\
&&= \frac{1}{Z} \Tr \left[\tilde{T}_{i_0, j_0, k_0, l_0,\lambda} \tilde{T}_{i_{\hat{\nu}}, j_{\hat{\nu}}, k_{\hat{\nu}}, l_{\hat{\nu}}, \lambda} T \cdots T\right].
\een

The number density is calculated by the numerical differentiation of the free energy in terms of $\mu$.
In case of $\mu_1\ne\mu_2$ the number density is defined as
\begin{equation}
	\la n_\lambda \ra = \frac{1}{V}\frac{\partial\ln Z}{\partial \mu_\lambda}=2N^2\beta\left\langle s_i(0) D^{(\mu_\lambda)}_{ij} s_j(\hat{2}) \right\rangle
\end{equation}
with $\lambda=1$ and 2. The case of $\mu_1=\mu_2=\mu$ is denoted by
\begin{equation}
	\la n \ra = \frac{1}{V}\frac{\partial\ln Z}{\partial \mu}=2N^2\beta\left\langle s_i(0) D^{(\mu)}_{ij} s_j(\hat{2}) \right\rangle,
\end{equation}
where $D^{(\mu_\lambda)}_{ij} \equiv \partial_{\mu_\lambda} D_{ij}(\mu_1,\mu_2,\hat{2})$ with
\begin{align}
	D^{(\mu_1)} =& \left(\begin{array}{cccc}
		\sinh(\mu_1) & ~0~ & ~0~ & -i\cosh(\mu_1) \\
		0 & 0 & 0 & 0 \\
		0 & 0 & 0 & 0 \\
		i\cosh(\mu_1) & 0 & 0 & \sinh(\mu_1)
	\end{array}\right), \\
	D^{(\mu_2)} =& \left(\begin{array}{cccc}
		~0~ & 0 & 0 & ~0~ \\
		0 & \sinh(\mu_2) & i\cosh(\mu_2) & 0  \\
		0 & -i\cosh(\mu_2) & \sinh(\mu_2) & 0 \\
		0 & 0 & 0 & 0 
	\end{array}\right), \\
	D^{(\mu)} =& \left(\begin{array}{cccc}
		\sinh(\mu) & 0 & 0 & -i\cosh(\mu) \\
		0 & \sinh(\mu) & i\cosh(\mu) & 0  \\
		0 & -i\cosh(\mu) & \sinh(\mu) & 0 \\
		i\cosh(\mu) & 0 & 0 & \sinh(\mu)
	\end{array}\right).
\end{align}


\section{Numerical results} 
\label{sec:results}


The partition function of Eq.~\eqref{eq:Z_TN} is evaluated using the HOTRG algorithm on lattices of the volume $V=L\times L$ with the periodic boundary condition in all the directions. 
In the following, all the results are calculated on the $L=1024$ lattice, where the TRG computation converges with respect to the system size and allows us to access the thermodynamic limit. Note that the correlation length of this model reaches $O(10^3)$ at $\beta=1.5$~\cite{Mana:1996nz}. In order to avoid contaminations from the finite size effects our results are restricted to $0< \beta \le 1.5$ in the following.

\begin{figure}[!ht]
	\centering
	\includegraphics[width=1.0\hsize]{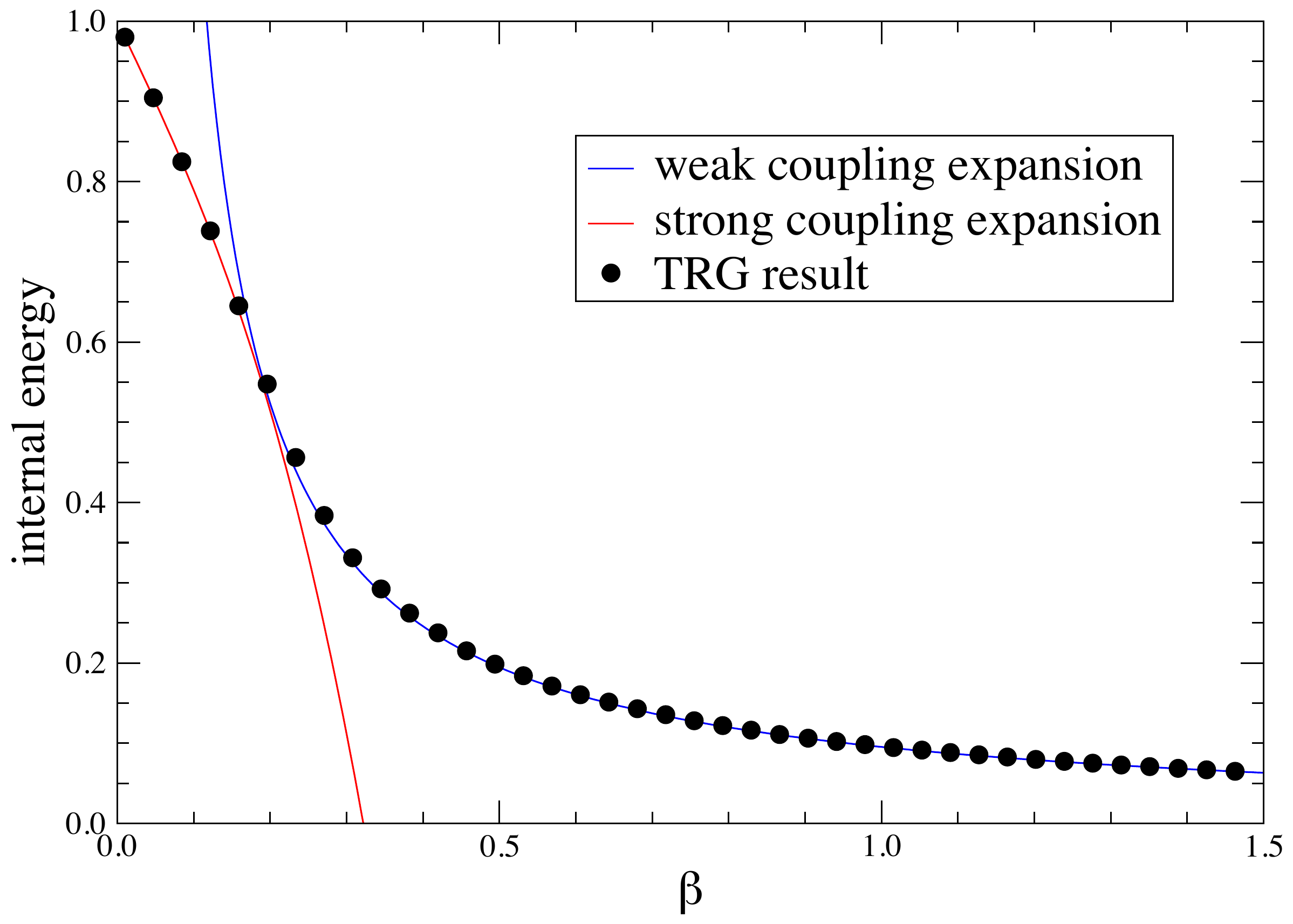}
	\caption{$\beta$ dependence of internal energy at $\mu=0$ on a lattice with $L=1024$. Solid curves denote the result of the strong and weak coupling expansions.}
  	\label{fig:internalenergy}
\end{figure}

\begin{figure}[!ht]
	\centering
	\includegraphics[width=1.0\hsize]{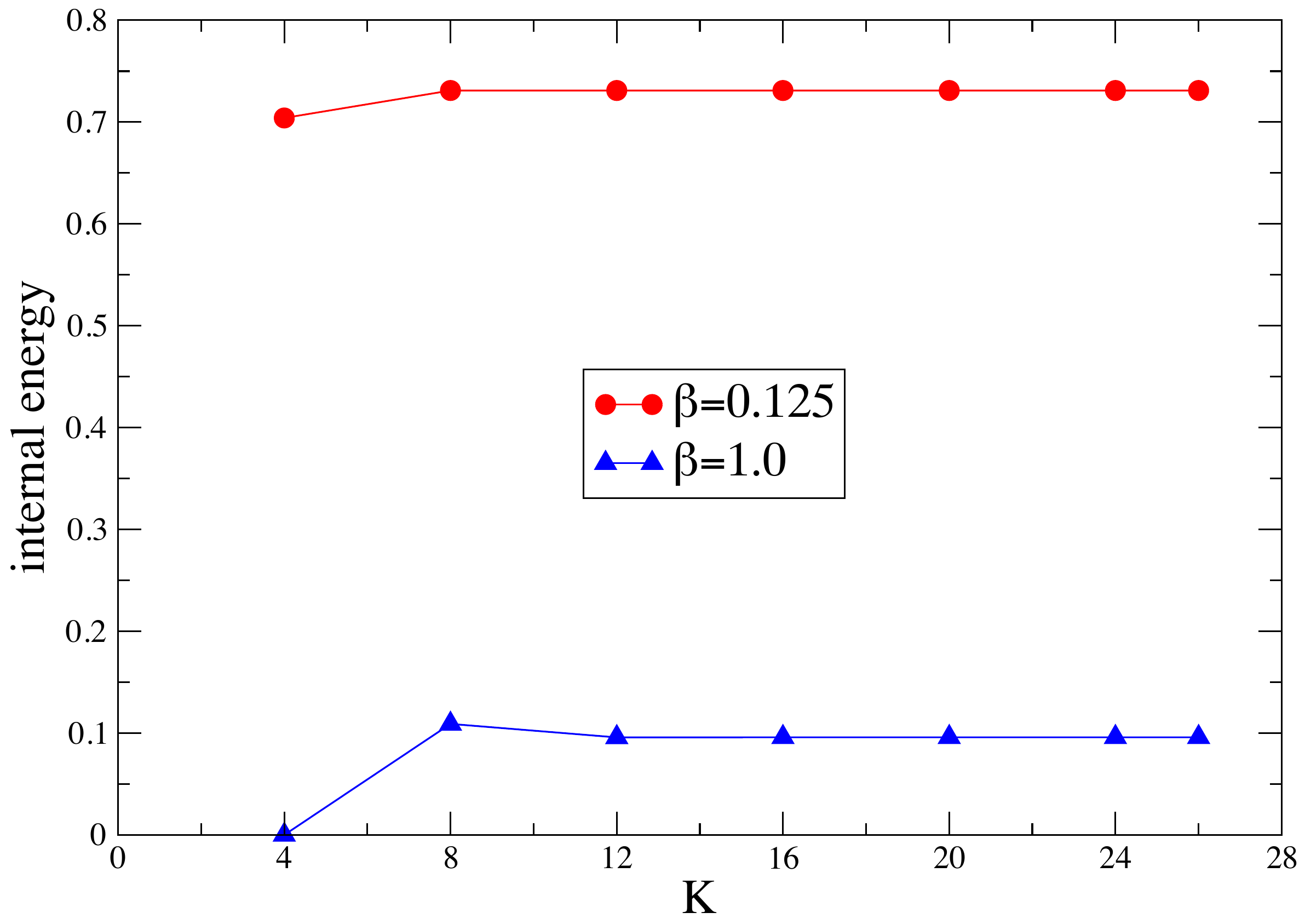}
	\caption{$K$ dependence of internal energy at $\mu=0$ with $D_\text{cut}=62$ on a lattice with $L=1024$.}
  	\label{fig:k-dep}
\end{figure}

\begin{figure}[!ht]
	\centering
	\includegraphics[width=1.0\hsize]{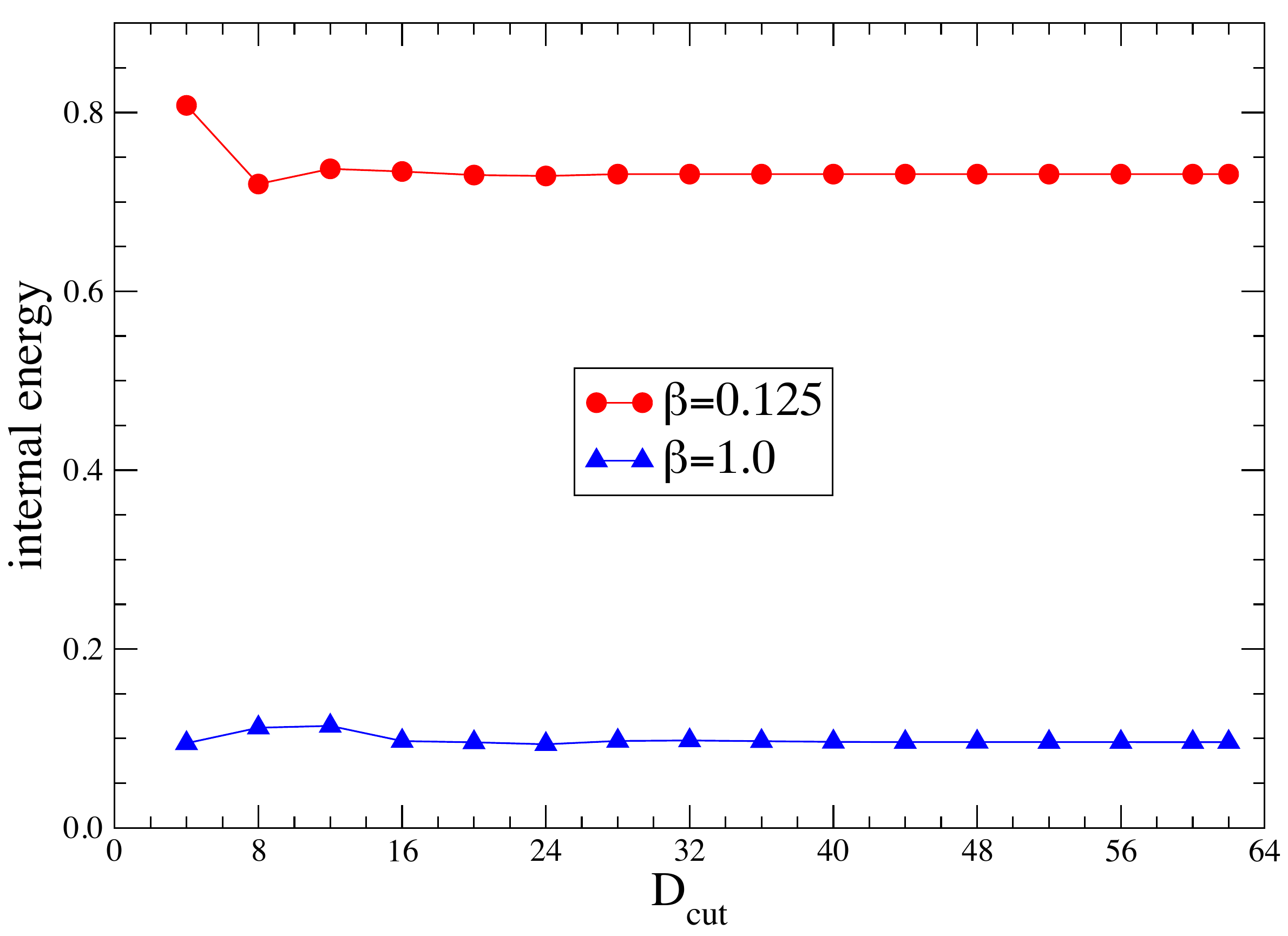}
	\caption{$D_\text{cut}$ dependence of internal energy at $\mu=0$ with $K=26$ on a lattice with $L=1024$.}
  	\label{fig:d-dep}
\end{figure}

We first show the results for the internal energy at $\mu=0$ in Fig.~\ref{fig:internalenergy}, which are obtained with the impure tensor method choosing $K=26$ and $D_\text{cut}=50$. In the strong coupling region we observe that our result shows good consistency with that of the strong coupling expansion up to $\beta\sim 0.15$. On the other hand, the result starts to follow the weak coupling expansion line around $\beta\sim 0.6$.   
Figures~\ref{fig:k-dep} and \ref{fig:d-dep} show the $K$- and the $D_\text{cut}$-dependence of the internal energy, respectively, choosing $\beta=0.125$ and 1.0 as representative $\beta$ values in the strong and weak coupling regions. 
The result converges as both $K$ and $D_\text{cut}$ increases and the combination of $K = 26$ and $D_\text{cut} = 50$,
which are fixed in the following calculation, are large enough to obtain converged results.

\begin{figure}[!ht]
	\centering
	\includegraphics[width=1.0\hsize]{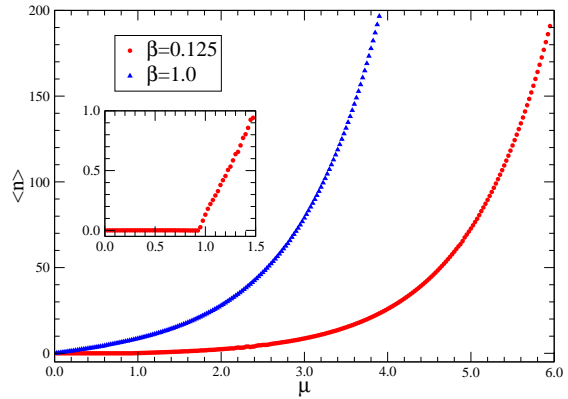}
	\caption{$\mu$ dependence of $\langle n\rangle$ at $\beta=0.125$ and 1.0 on a lattice with $L=1024$. Inset graph magnifies the result in small $\mu$ region for $\beta=0.125$.}
  	\label{fig:numberdensity_n}
\end{figure}

\begin{figure}[!ht]
	\centering
	\includegraphics[width=1.0\hsize]{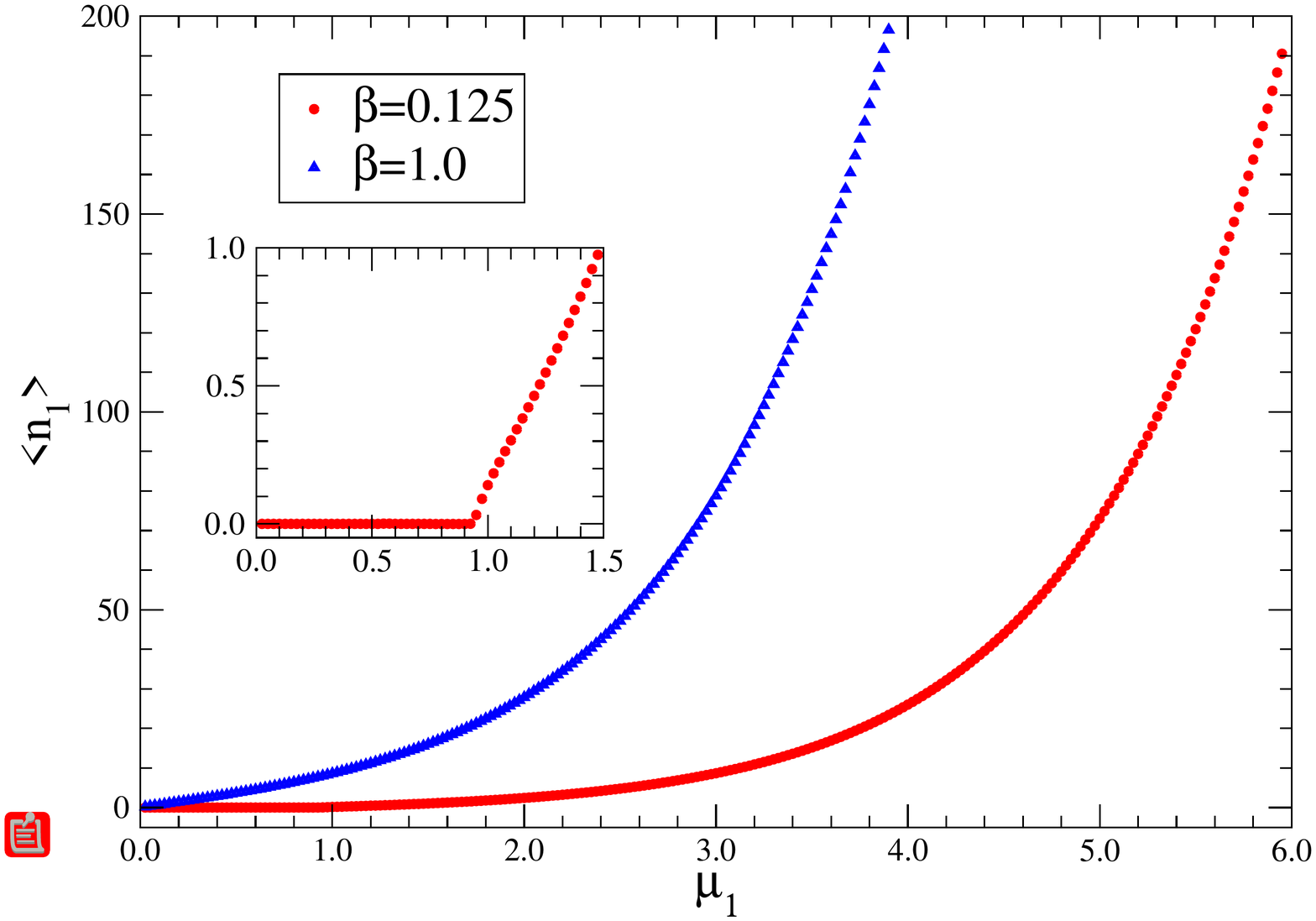}
	\caption{$\mu_1$ dependence of $\langle n_1\rangle$ with $\mu_2=0$ at $\beta=0.125$ and 1.0 on a lattice with $L=1024$. Inset graph magnifies the result in small $\mu$ region for $\beta=0.125$.}
  	\label{fig:numberdensity_n1}
\end{figure}

\begin{figure}[!ht]
	\centering
	\includegraphics[width=1.0\hsize]{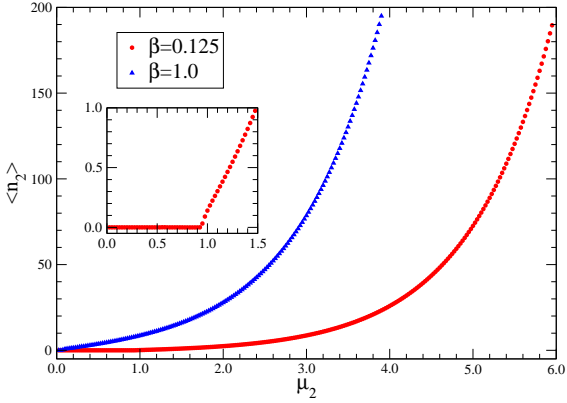}
	\caption{$\mu_2$ dependence of $\langle n_2\rangle$ with $\mu_1=0$ at $\beta=0.125$ and 1.0 on a lattice with $L=1024$. Inset graph magnifies the result in small $\mu$ region for $\beta=0.125$.}
	\label{fig:numberdensity_n2}
\end{figure}

Let us turn to the finite density case, where
an interesting physical quantity is the number density.
We evaluate it with the numerical differentiation of the free energy in terms of $\mu$ as explained in Sec.~\ref{subsec:tn-rep}. Figure~\ref{fig:numberdensity_n} shows the $\mu$ dependence of $\langle n\rangle$ at $\beta=0.125$ and $1.0$. The former is a representative case in the strong coupling region and the latter belongs to the weak coupling region. Although both results seem to increase monotonically from zero as $\mu$ increases in the large scale, we find a clear Silver-Blaze phenomena in the small $\mu$ region at $\beta=0.125$ in the inset. The similar $\beta$ dependence of the number density is found in the 4$d$ case~\cite{Gattringer:2017hhn}, where the Silver-Blaze region becomes narrower as $\beta$ increases.
We also plot $\mu_1$ dependence of $\langle n_1\rangle$ with $\mu_2=0$ at $\beta=0.125$ and $1.0$ in Fig.~\ref{fig:numberdensity_n1}.  The $\mu_1$ dependence of $\langle n_1\rangle$ is quite similar to the $\mu$ dependence of $\langle n\rangle$ in Fig.~\ref{fig:numberdensity_n}. Figure~\ref{fig:numberdensity_n2} shows the $\mu_2$ dependence of $\langle n_2\rangle$ with $\mu_1=0$, whose behavior is also similar to that of $\langle n\rangle$.

\section{Summary and outlook} 
\label{sec:summary}

In this work we have shown the efficiency of the Gauss quadrature to discretize the SU(2) Lie group using the (1+1)$d$ SU(2) PCM at finite density. The internal energy at $\mu=0$ shows good consistency with the predictions of the strong and weak coupling expansions over $0<\beta\le 1.5$. We have successfully evaluated the number density in the finite density region, which shows the Silver-Blaze phenomena in the strong coupling region. The next step would be to study the higher-dimensional SU(2) gauge theory with the Gauss quadrature method.

\begin{acknowledgments}
  Numerical calculation for the present work was carried out with the supercomputer Cygnus under the Multidisciplinary Cooperative Research Program of Center for Computational Sciences, University of Tsukuba.
This work is supported in part by Grants-in-Aid for Scientific Research from the Ministry of Education, Culture, Sports, Science and Technology (MEXT) (No. 20H00148).

\appendix

\section{(1+1)$d$ O(3) nonlinear sigma model at finite density}

In order to check the validity of the TRG method with the Gauss-Legendre quadrature we apply it to the  (1+1)$d$ O(3) nonlinear sigma model at finite density and compare the results with those obtained by the dual lattice simulation~\cite{Bruckmann:2016txt}. The action is defined by
\begin{equation}
	S = -\beta \sum_{n\in\Lambda_{1+1},\nu} s_\lambda(\theta_n,\phi_n) D_{\lambda\gamma}(\mu,\hat{\nu}) s_\gamma(\theta_{n+\hat{\nu}},\phi_{n+\hat{\nu}}), 
\end{equation}
where spin $s_\lambda(\theta,\phi)$ and matrix $D_{\lambda\gamma}(\mu,\hat{\nu})$ are 
\begin{align}
	& s(\theta_{a_n},\phi_{b_n})= \left(\begin{array}{l}\cos\theta_{a_n}\\
		\sin\theta_{a_n}\cos\phi_{b_n}\\
		\sin\theta_{a_n}\sin\phi_{b_n}
	\end{array}\right) \\
	& D(\mu,\hat{\nu}) = \left(\begin{array}{rrr} 1 & ~ & ~ \\
		~ & \cosh(\delta_{2,\nu}\mu) & -i\sinh(\delta_{2,\nu}\mu) \\
		~ & i\sinh(\delta_{2,\nu}\mu )& \cosh(\delta_{2,\nu}\mu)
	\end{array}\right) 
\end{align}
respectively. $\theta_{a_n}$, $\phi_{b_n}$ are $a$-, $b$-th roots of the $K$-th Legendre polynomial $P_{K}(s)$ on the site $n$, respectively. The tensor network representation of (1+1)$d$ O(3) nonlinear sigma model with Gauss-Legendre quadrature is similar to SU(2) PCM
\begin{widetext}
\begin{align}
	\label{eq:O3puretensor}
	T_{i_n, j_n, k_n, l_n} &= \frac{\pi}{8} \sqrt{\sigma_{i_n}(1) \sigma_{j_n}(1) \sigma_{k_n}(2) \sigma_{l_n}(2) } \sum_{a_n, b_n, c_n} w_{a_n} w_{b_n} \sin(\theta_{a_n})\nonumber \\
	&\quad \times V^\dagger_{i_n,a_n,b_n} (1) U_{a_n,b_n, j_n} (1) V^\dagger_{k_n,a_n,b_n} (2) U_{a_n,b_n, l_n} (2)\\
	\label{eq:O3internalenergytensor1}
	\tilde{T}_{i_0, j_0, k_0, l_0,\lambda} &= \frac{\pi}{8} \sqrt{\sigma_{i_0}(1) \sigma_{j_0}(1) \sigma_{k_0}(2) \sigma_{l_0}(2) } \sum_{a_0, b_0,} w_{a_0} w_{b_0} \sin(\theta_{a_0})  \nonumber \\
	&\quad \times s_\lambda(a_0,b_0) V^\dagger_{i_0,a_0,b_0} (1) U_{a_0,b_0, j_0} (1) V^\dagger_{k_0,a_0,b_0} (2) U_{a_0,b_0, l_0} (2), \\
	\label{eq:O3internalenergytensor2}
	\tilde{T}_{i_{\hat{\nu}}, j_{\hat{\nu}}, k_{\hat{\nu}}, l_{\hat{\nu}}, \lambda} &= \frac{\pi}{8} \sqrt{\sigma_{i_{\hat{\nu}}}(1) \sigma_{j_{\hat{\nu}}}(1) \sigma_{k_{\hat{\nu}}}(2) \sigma_{l_{\hat{\nu}}}(2) } \sum_{a_{\hat{\nu}}, b_{\hat{\nu}}, c_{\hat{\nu}}} w_{a_{\hat{\nu}}} w_{b_{\hat{\nu}}} \sin(\theta_{a_{\hat{\nu}}}) \nonumber\\ 
	&\quad \times D_{\lambda\gamma}(\mu,\hat{\nu}) s_\gamma(a_{\hat{\nu}},b_{\hat{\nu}})  V^\dagger_{i_{\hat{\nu}},a_{\hat{\nu}},b_{\hat{\nu}}} (1) U_{a_{\hat{\nu}},b_{\hat{\nu}}, j_{\hat{\nu}}} (1) V^\dagger_{k_{\hat{\nu}},a_{\hat{\nu}},b_{\hat{\nu}}} (2) U_{a_{\hat{\nu}},b_{\hat{\nu}}, l_{\hat{\nu}}} (2) ,
\end{align}
\end{widetext}
The number density can be obtained by replacing the matrix $D$ with
\begin{equation}
	D^{(\mu)} = \left(\begin{array}{rrr} 0 & ~ & ~ \\
		~ & \sinh(\mu) & -i\cosh(\mu) \\
		~ & i\cosh(\mu )& \sinh(\mu)
	\end{array}\right) .
\end{equation}

Figure~\ref{fig:o3_energy} shows the results for the internal energy at $\mu=0$ obtained with the impure tensor method choosing $K=100$ and $D_{\rm cut}=48$. As in the SU(2) PCM case, our results show good consistency with the strong and weak coupling expansions in the strong and weak coupling regions, respectively. 
In Fig.~\ref{fig:o3_n} we present the results for the number density at $\beta=1.0$, 1.2 and 1.4. We observe that the Silver Blaze behavior beomes sharper as the lattice size increases from $L=64$ to $L=1024$. In comparison with Fig.~2 in Ref.~\cite{Bruckmann:2016txt}, the dual lattice simulation on a $90\times 90$ lattice shows better consisitency with our results with $L=64$ than those with $L=1024$.

\begin{figure}[!ht]
	\centering
	\includegraphics[width=1.0\hsize]{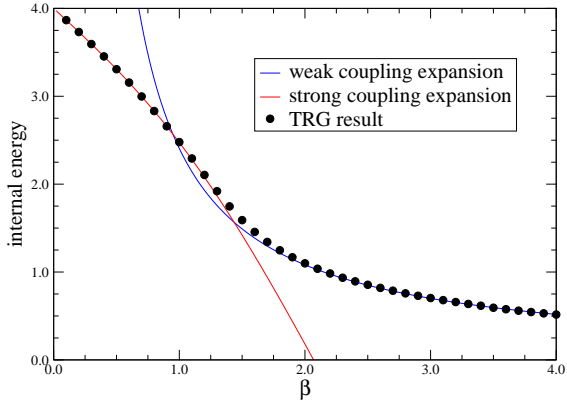}
	\caption{$\beta$ dependence of internal energy at $\mu=0$ for the (1+1)$d$ O(3) nonlinear sigma model at finite density. The lattice size is $L=1024$. Solid curves denote the results of the strong and weak coupling expansions.}
  	\label{fig:o3_energy}
\end{figure}

\begin{figure}[!ht]
	\centering
	\includegraphics[width=1.0\hsize]{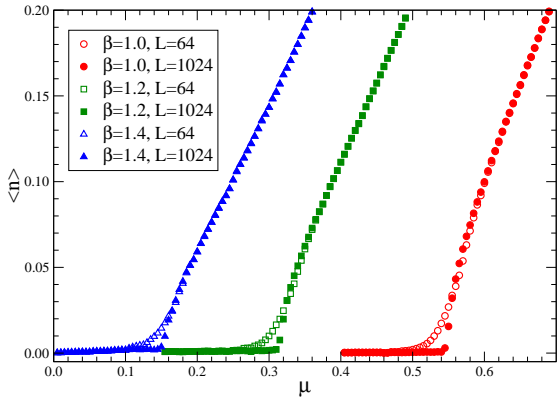}
	\caption{$\mu$ dependence of $\langle n\rangle$ for the (1+1)$d$ O(3) nonlinear sigma model at finite density. $\beta=1.0$, 1.2 and 1.4 are chosen on a lattice with $L=64$ (open) and 1024 (closed).}
  	\label{fig:o3_n}
\end{figure}

\end{acknowledgments}




\bibliographystyle{apsrev4-2}
\bibliography{formulation,algorithm,discrete,grassmann,continuous,gauge,review,for_this_paper}

\end{document}